\documentclass[journal,twoside,web]{ieeecolor}
\usepackage{soul}
\usepackage{tikz}
\usepackage{jsen}
\usepackage{cite}
\usepackage{amsmath,amssymb,amsfonts}
\usepackage{algorithmic}
\usepackage{graphicx}
\usepackage{textcomp}
\usepackage{wrapfig}
\def\BibTeX{{\rm B\kern-.05em{\sc i\kern-.025em b}\kern-.08em
    T\kern-.1667em\lower.7ex\hbox{E}\kern-.125emX}}
\definecolor{abstractbg}{rgb}{0.89804,0.94510,0.83137}
\newcommand\copyrighttext{%
  \footnotesize This work has been submitted to the IEEE for possible publication. Copyright may be transferred without notice, after which this version may no longer be accessible.}
\newcommand\copyrightnotice{%
\begin{tikzpicture}[remember picture,overlay]
\node[anchor=south,yshift=10pt] at (current page.south) {\fbox{\parbox{\dimexpr\textwidth-\fboxsep-\fboxrule\relax}{\copyrighttext}}};
\end{tikzpicture}%
}

\begin{document}
\title{A plug-and-play type field-deployable bio-agent free salicylic acid sensing system}
\author{Bhuwan Kashyap, \IEEEmembership{Member, IEEE}, and Ratnesh Kumar, \IEEEmembership{Fellow, IEEE}
\thanks{This work was supported in part by the National Science Foundation under the grants CCF-1331390, ECCS-1509420, PFI-1602089, and CSSI-2004766.}
\thanks{Bhuwan Kashyap, PhD scholar in the Department of Electrical and Computer Engineering, Iowa State University, Ames, IA 50010 USA (e-mail: bkashyap@iastate.edu).}
\thanks{Ratnesh Kumar, Harpole Professor in the Department of Electrical and Computer Engineering, Iowa State University, Ames, IA 50010 USA (e-mail: rkumar@iastate.edu)}}

\IEEEtitleabstractindextext{%
\fcolorbox{abstractbg}{abstractbg}{%
\begin{minipage}{\textwidth}%
\begin{wrapfigure}[17]{r}{3in}%
\includegraphics[width=3in]{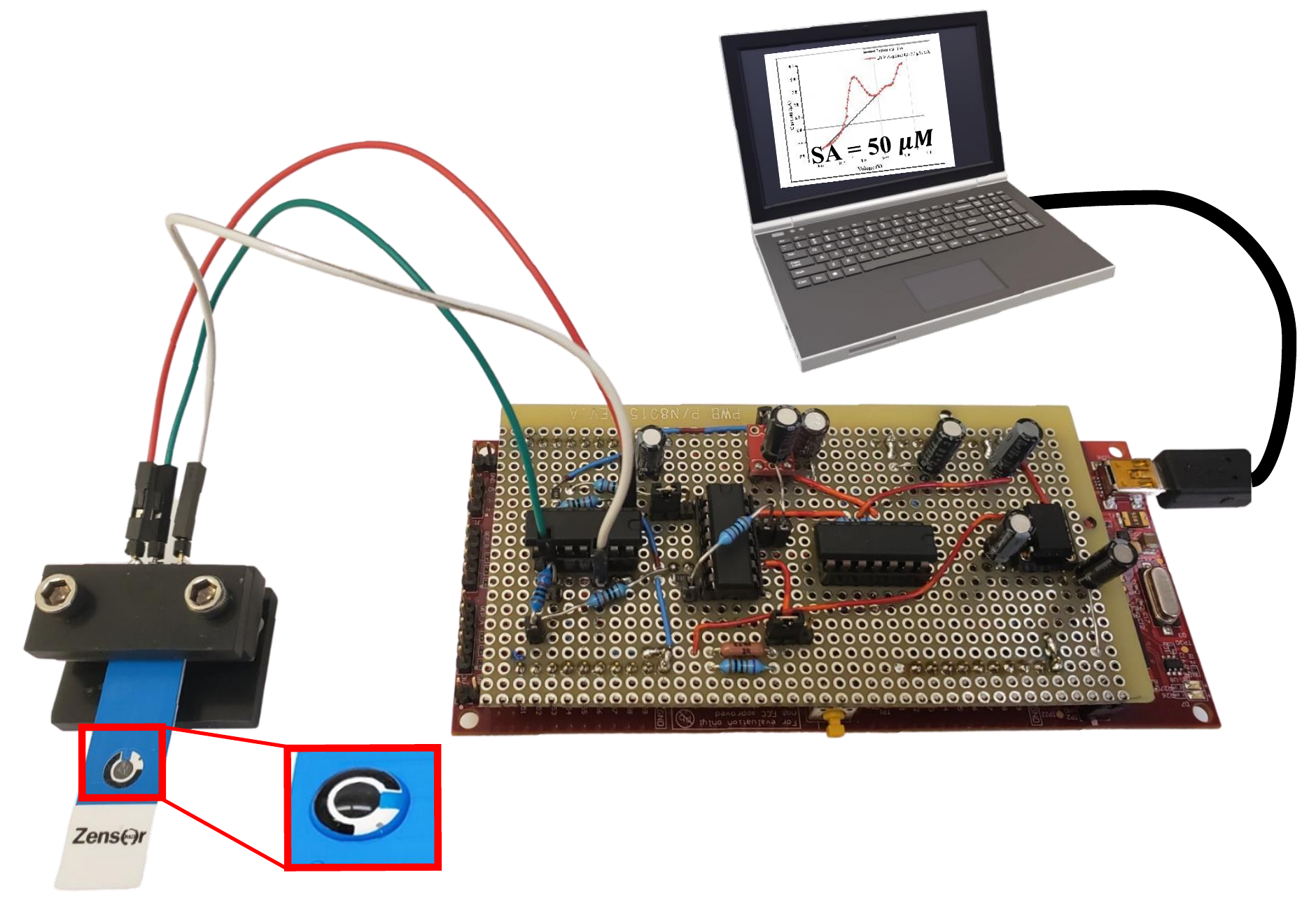}%
\end{wrapfigure}%
\begin{abstract}

Salicylic acid (SA) is a primary phytohormone released in response to stress, particularly biotic infections in plants. Monitoring SA levels may provide a way for early disease detection in crops providing a way for applying effective measures for reducing agricultural losses while increase our agricultural efficiency. Additionally, SA is an important chemical used extensively in the pharmaceutical and healthcare industry due to its analgesic and anti-inflammatory properties. Developing a fast and accurate way for monitoring SA levels in human serum can have a life-saving impact for patients suffering from overdosing and/or mis-dosing. In this work, we present a low-cost, portable, and field-deployable electrochemical SA sensing system aimed towards achieving the above-mentioned goals. The developed sensor consists of a plug-and-play type device equipped with specialized designed high accuracy sensing electronics and a novel procedure for robust data analysis. The developed sensor exhibits excellent linearity and sensitivity and selectivity. The practical applicability of the developed sensor was also demonstrated by measuring SA levels in real samples with good accuracy.

\end{abstract}

\end{minipage}}}

\maketitle
\copyrightnotice

\section{Introduction}
\label{sec:introduction}
\IEEEPARstart{T}{here} is an ever increasing demand for higher crop production to keep up with the needs of the growing human population which is expected to reach 9.8 billion in 2050 and 11.2 billion in 2100 \cite{intro_1}. Achieving the required agricultural output sustainably depends directly on optimal nutrition and water application while reducing yield losses, thereby improving efficiency in crop production. Presently, the use of synthetic pesticides, bactericides and fungicides enables growers to improve crop yields which comes at the expense of environmental contamination as many of these chemicals are toxic in nature. Besides having an environmental and economical impact, the overuse of these chemicals in soil may lead to pathogen resistance as well as have long term effects on soil health.

Plants are constantly exposed to the elements of nature, facing countless biotic and abiotic stressors throughout their lifetime. One way plant achieve immunity is through local hypersensitive response at the attack site that can immunize the plant against further attack, this phenomenon was termed as \emph{systemic acquired resistance} (SAR) by A. Frank Ross in 1961 \cite{SA_role0}. The onset of SAR is accompanied by increased accumulation of signalling hormones at the attack site and their transport to other plant tissues via the phloem \cite{SA_role3}. 2-hydroxybenzoic acid or salicylic acid (SA) is one such key signalling phytohormone responsible for the activation of SAR against biotic stresses, primarily due to biotrophic pathogens \cite{SA_role2, SA_role3}. Other major signalling hormones include Jasmonic acid (JA), Ethylene (ET) and Abscisic acid (ABA) where, JA is associated with defense against herbivorous insects and necrotrophic pathogens whose pathways are modulated by ET and, ABA is associated with abiotic stress responses. Overall, SA and JA are the main phytohormones activated during SAR while, ET, ABA and other secondary metabolites play more modulating roles \cite{SA_role4, SA_role1}. In general, a complex crosstalk between the SA and JA pathways exist providing robust immune signalling. The study and exploration of these mechanisms is currently an active area of research \cite{SA_role1, SA_role6}.

While endogenous accumulation of SA in plant tissue has been largely associated with pathogenic stresses, SA has also been studied to have an effect on abiotic stresses \cite{SA_role12}. Exogenous application of SA and related compounds have been reported to increase tolerance in plants against various abiotic stress such as salinity, radiation, chilling, heat and more \cite{SA_role5, SA_role7, SA_role8, SA_role1}.

Therefore, developing technologies for rapid measurement of phytohormones (in this case SA) can facilitate early detection of plant stress levels resulting in efficient and timely responses, minimizing yield losses. In addition to having a direct impact on predictive plant health monitoring, hormone sensing can help us improve our understanding of the growth and immune signalling processes, paving way for our ability to manipulate and control SAR in crop species that would provide farmers with effective and environmentally friendly ways to prevent yield losses.

Besides its significance in the agriculture, SA is also of importance in other industries such as, healthcare, pharmaceutical and cosmetics due to its analgesic, antiseptic and anti-inflammatory properties. Salicylates are an active ingredient in many over-the-counter non steroidal anti-inflammatory drugs like Aspirin, which are among the most commonly used medication for treating acute as well as chronic pains worldwide. However, their overdose and/or accidental misuse/mis-dosing may lead to salicylate poisoning, and according to the American Association of Poison Control Centers (AAPCC), 24 \% of analgesic-related deaths can be attributed to aspirin (alone or in combination with other drugs) \cite{intro_2}. Moreover, a small percentage of the population suffers from salicylate sensitivity which can result in gastrointestinal issues when foods high in salicylates are consumed \cite{intro_3, intro_4}. Therefore, fast and early detection of SA levels in human serum and urine samples can have a potential life saving impact. 

In this work, a plug-and-play-type bio-agent free portable SA sensing system is presented. The working principle of the proposed sensor involves electrochemical (EC) characterization of electro-oxidation of SA on a carbon electrode using differential pulse voltammetry (DPV) technique. The overall system consists of the following main constituents: 

\begin{enumerate}
\item Plug-and-play system compatible with the screen printed carbon electrode (SPCE).
\item Application specific sensor electronics.
\item Data analytics for robust DPV signal processing.
\end{enumerate}

The sensor presented in this work is the \emph{first-of-its-kind portable plug-and-play type device} developed for SA sensing. Our sensor exhibits excellent response providing a novel, economical, accurate and portable detection system for SA where the total cost of a single unit of the developed SA sensor is under \$50. This paper has been divided into the six sections including introduction, related works, materials and methods, portable SA sensing system, results and discussion, and conclusion.

\begin{figure}[h]
\centering
\includegraphics[width=0.45\textwidth]{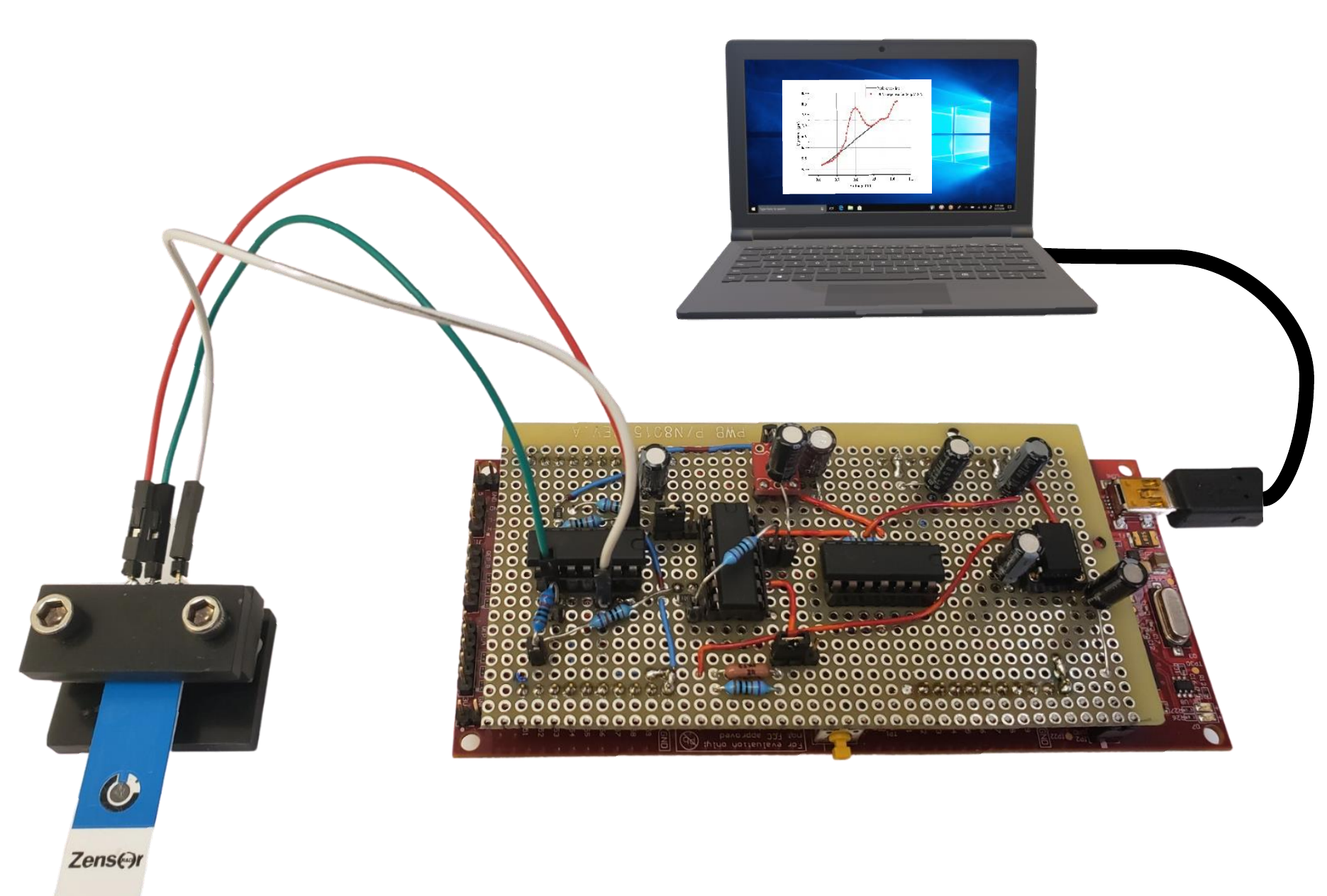} 
\caption{The overall portable SA sensing system}
\label{complete}
\end{figure}

\section{Related works} \label{literature}

Quantitative determination of SA has potential applications spanning several industries, such as agriculture, food, pharmaceutical and healthcare, for which various sensing approaches have been explored in literature. Among analytical techniques, methods based on high performance liquid chromatography (HPLC) \cite{SA_sens1, SA_sens2}, and mass spectrometry (MS) or chromatography coupled tandem mass spectrometry \cite{SA_sens3, SA_sens4, SA_sens5, SA_sens6} are routinely used for detecting SA. Analytical methods offer high accuracy and precision but are limited to laboratory setting as they require expensive sophisticated equipment (non-portable), extensive sample preparations and training, and are time, cost and labor intensive. Based on the need for quick response, minimal sample preparation, ease of use and potential for \emph{in-situ} operation or real-time analysis, various biosensing (involving bio-molecules such as enzymes, aptamers and molecularly imprinted polymers)and electro-analytical reactions of SA based approaches are continuously being developed. 

Some of the key recent advances and related works in biosensing for SA are discussed here. One such method involves detection of SA using a bi-enzyme system consisting of salicyate hydroxylase (SH) and tyrosinase (TYR). SH catalyzes the conversion of SA in to catechol, and then, TYR catalyzes catechol to quinone. The formed quinone can produce back catechol and this recurrent conversion between involves an electron transfer which can be characterized using EC methods. An EC sensor fabricated using carbon nano-tubes (CNTs) modified carbon electrode employing the aforementioned bi-enzyme recipe is reported in \cite{SA_sens10}. The sensor was characterized using cyclic voltammetry (CV) and constant potential amperometry exhibiting sensitivity as 30 $\mu A cm^{-2} \mu M^{-1}$ and limit of detection (LOD) as 13 $nM$. In another work, a bi-enzymatic microfluidic EC sensing device was reported with SA detection range of 0.5 $\mu M$ to 64 $\mu M$ characterized using chrono-amperometry \cite{SA_sens8}. Alternatively,molecularly imprinted polymers (MIPs), which are a artificial antibodies prepared by polymerizing functional and cross-linking monomers around specific target molecules, have also been reported for SA detection. A dual functional MIP-modified organometal lead halide perovskite biosensor was reported for photoelectrochemical bioanalysis of SA in \cite{SA_sens7}. The binding between the MIP and SA due to shape and hydrogen bond recognition affects the recorded photocurrent which was calibrated with SA concentration. The sensor indicated logarithmic response with LOD of $1.95 \times 10^{-13} M$, but with limited range of operation with the upper bound of $1 \times 10^{-8} M$ which may be unsuitable for agricultural applications. In a different recent work, MIP functionalized $TiO_2$ nanorod arrays were described for SA recognition and detection \cite{SA_sens11}. The sensor was reported with LOD of $3.9 \times 10^{-8} M$ and range of $1.0 \times 10^{-7} M$ to $5.0 \times 10^{-5} M$. SA sensing has also been realized using aptamers which are single-stranded DNA, or RNA molecules that can be made with high affinity for desired analyte. An aptamer-based label free SA nanosensor using structure-switching systemic evolution of ligands by exponential enrichment (SELEX) technique was developed in \cite{SA_sens9}. The identified SA aptamer was incorporated onto a nanostructured Fabry-Perot interference sensor where the interference fringes were used as transducing signals for SA quantification down to 0.1 $\mu M$. Sensing strategies utilizing bio-molecules for SA detection are actively being developed, and while they offer sensitive detection with good selectivity for relatively rapid response and potential for \emph{in-situ} operation, they also exhibit some limitations. Firstly, bio-molecules require elaborate, costly and complicated manufacturing processes, and their characteristics depend heavily on the method of synthesis imparting inherent variability in operation. Secondly, biosensors involve complex, expensive and time consuming fabrication procedures, such as functionalization of bio-molecules on electrode surfaces and transduction mechanisms, which are prone to inconsistencies in performance. Finally, the sensors have relatively short shelf life while needing special storage and operating conditions as bio-molecules are susceptible to leaching and/or getting denatured.  

Besides biosensing strategies, electro-catalytic reactions on metallic and carbon electrodes have also been characterized for SA quantification. Cerium-doped zirconium oxide (Ce/ZrO$_2$) was introduced as an electrocatalyst for the electro-oxidation of SA in a recent work, and Square wave voltammetry was used to quantify the analytical signal in the concentration range of 5 $\mu$M to 1000 $\mu$M with LOD of 1.1 $\mu$M\cite{SA_role9}. In another work, a paper-based electro-analytical device for in-situ determination of SA in tomato leaves was reported. DPV was used to signalize from 0.5 $\mu$M to 100 $\mu$M with moderate accuracy, and selectivity was tested with common interfering species in plant sap \cite{SA_sens15}. Other electrode materials used for electro-oxidation based analysis of SA include, electro-reduced graphene oxide modified screen printed carbon electrode (ERGO-SPCE)\cite{SA_sens12}, modified electrode. well-aligned multiwalled carbon nanotube electrode \cite{SA_sens17}, screen printed graphite electrode \cite{SA_sens14}, carbon-fiber electrode \cite{SA_sens16}, and carbon nanotube/iron oxide nanoparticle (SWCNT/ION) modified electrode \cite{SA_sens13}. SA electro-oxidation mechanism and its effect on the electrode surface leading to fouling caused by an electro-polymerized passive SA-film was explored in \cite{SA_sens16}.It was postulated that most previously reported SA sensor responses may have been misinterpreted. 


\section{Materials and methods}

The materials used and their sources, and describes the SA sensing methodology applied in this work.

\subsection{Materials used} \label{material}

SA (powdered), sodium hydroxide (NaOH), potassium ferricyanide (K$_3$[Fe(CN)$_6$]), glucose, indole-3-acetic acid (IAA) and phospahte buffered saline (PBS) of pH 6.6 buffer was purchased from Millipore-Sigma, MO, USA. Ethanol (200 Proof), potasssium chloride (KCl), methyl jasmonate (MeJA), abscisic acid (ABA), citric acid, uric acid, succinic acid, and malic acid was purchased from Thermo Fisher Scientific, MA, USA. Disposable screen printed carbon electrodes (Zensor brand) and the bench-top potentiostat CHI660E were purchased from CHI instruments Inc., TX, USA. Salicylate liqui-UV test kit was purchased from EKF diagnostics USA (Stanbio labs), TX, USA, and was used with UNICO SQ2800 UV/VIS spectro-photometer. Deionized (DI) water with a conductivity of 18.2 M$\Omega$ was used to prepare all the solutions. Texas instrument's (TI) C2000 Launchxl-F28379D (MCU Launchpad) and other electrical components (listed in the Appendix A) were purchased from DigiKey Electronics, MN, USA. The MCU was programmed using Code composer studio v10 (CCS v10), available free of cost at TI.com, while Matlab 2019a was used for data analytics. 

\subsection{SA sensing methodology}

\begin{figure}[h]
\centering
\includegraphics[width=0.48\textwidth]{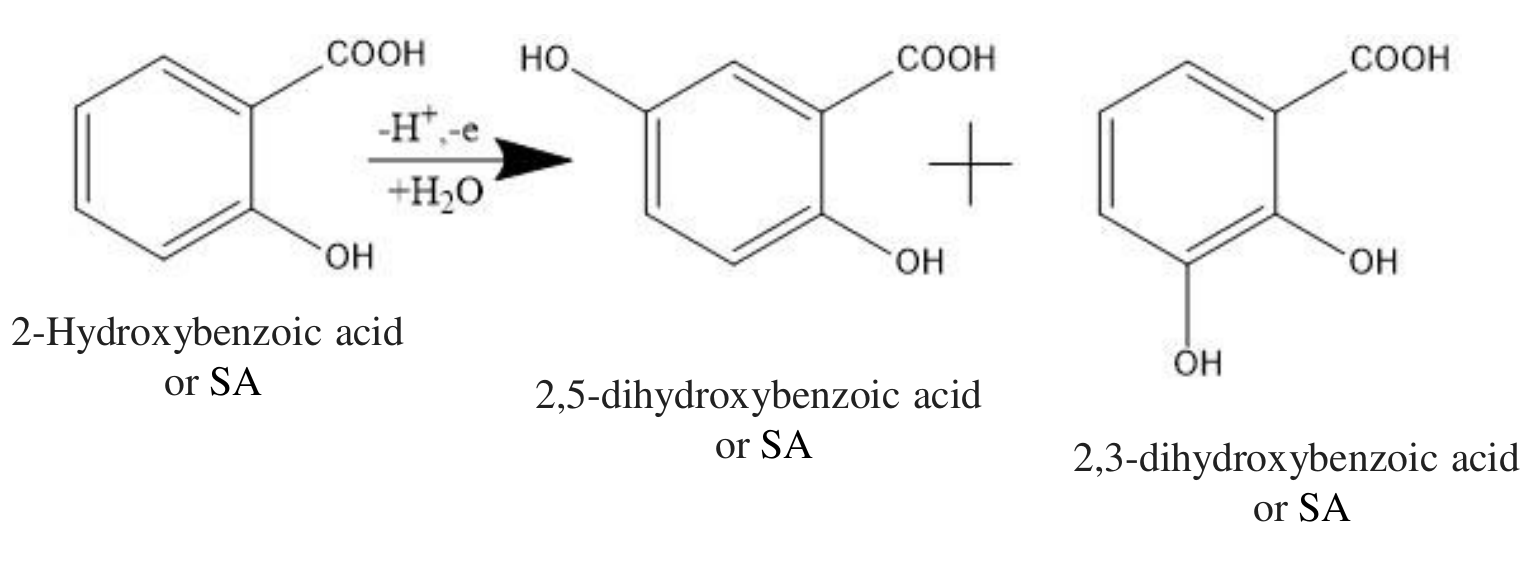} 
\caption{Schematic of the developed portable SA sensing system. (A) Input signal conditioning circuit; (B) Core potentiostat circuit; (C) Output signal detecting circuit; (D) Single-ended-to-differential converter circuit}
\label{method}
\end{figure}

In this work, electro-oxidation-based sensing methodology was applied towards SA detection, where the primary chemical reaction is shown in the figure \ref{method}. This method is especially suitable for in-field SA testing applications as it does not involve the use of any bio-agents such as enzymes and/or MIPs that are prone to fouling, need specific operating conditions (in term of temperature and humidity), are complex and expensive to manufacture, exhibit large inter-sensor variability and have poor shelf-life. On the other hand, electrochemically characterizing the oxidation of SA on a carbon electrode as a way to selectively quantify the concentration of SA, not only overcomes the challenges presented by other methods but also provide a reliable way to develop field-deployable and cost-effective SA sensors. DPV was adopted as the electrochemical method of choice for the characterization of SA-oxidation in this work due its high sensitivity and good selectivity. The specificity in DPV-based sensing method comes from the fact that current-potential response (or the potential at which the current peak occurs) is specific to a particular chemical reaction under given ambient conditions. 

Furthermore, based on the reports published in literature \cite{SA_sens14, SA_sens16}, in addition to the primary products of the SA oxidation reaction (shown in figure \ref{method}) other secondary products may also be formed. It was determined that the secondary products form polymeric-SA (poly-SA) films that may passivate the electrode surface increasing the electrode resistance and therefore reducing the current response. Following on the results reported in \cite{SA_sens16}, the screen printed carbon electrodes were immersed in 0.1 M NaOH solution for five minutes before each measurement in order to remove such passivation layers which may lead to improved linearity in DPV response.

\section{Portable SA sensing system}

\begin{figure}[h]
\centering
\includegraphics[width=0.49\textwidth]{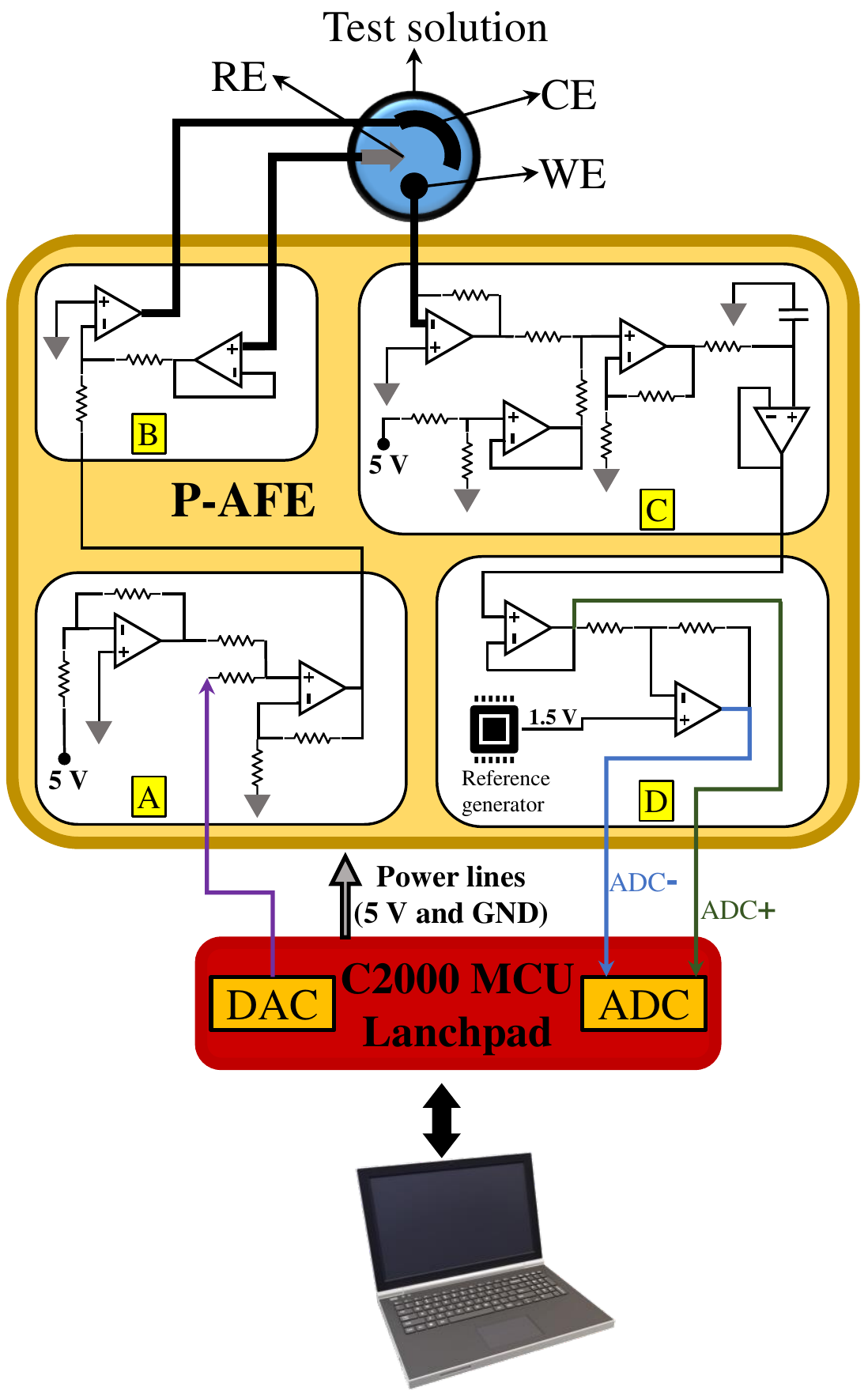} 
\caption{Schematic of the developed portable SA sensing system. (A) Input signal conditioning circuit; (B) Core potentiostat circuit; (C) Output signal detecting circuit; (D) Single-ended-to-differential converter circuit}
\label{system}
\end{figure}

Figure \ref{system} presents the schematic of the complete field-deployable EC sensing system developed in this work, and its detailed description is provided in this section. The main components of the SA sensing system include a low-cost disposable electrode, the application-specific sensing electronics and the sensor data analysis. 

A commercially available screen-printed sensing electrode system (as shown in figures \ref{complete} and \ref{system}) comprising of a planar carbon working electrode (WE) integrated with a carbon counter electrode (CE) and an Ag/AgCl reference electrode (RE), arranged in a three-electrode configuration, was used in this work. The commercial electrodes offer reliable operation while being cost-effective making them ideal for developing a low-cost portable EC sensors. In order to connect the 3-electrode system with the sensing electronics, a specially designed adapter (shown in figure \ref{complete}) was fabricated using a 3D polymer printer (using B9 creator v1.2 3D printer). The custom-developed adapter allowed for an easy and reliable electrical interface between the sensing electronics and the employed planar screen-printed electrodes. 

The sensing electronics consisted of an application-specific portable potentiostat made of two main components: (i) TI's C2000-Launchpad microcontroller (MCU) and, (ii) the custom developed potentiostat analog-front-end (P-AFE) as shown in the figure \ref{system}. The P-AFE module was fabricated such that it can be affixed atop the MCU platform forming a compact portable instrument (refer figure~\ref{complete}). The overall circuit was specifically optimized for electro-oxidation-based SA detection using a three-electrode system, characterized by DPV. 

The overall circuit system consists of the following main sub-circuits; the input signal conditioning, the three-electrode potentiostat system, the output current-to-voltage converter and voltage level adjuster, and the single-ended to differential signal converter required for sampling the signal using the differential 16 bit ADC integrated in the C2000 MCU platform. 

According to the proposed architecture, first, the input DPV signal is generated using the integrated 12 bit digital-to-analog converter (DAC) on-board the MCU followed by the DC offset adjustment (as desired) on the P-AFE using a voltage adder circuit. The final DPV input signal is then applied to the core potentiostat sub-circuit which was developed based on a \emph{potential control amplifier in adder configuration} as described in \cite{book1}. In the next stage, the output current from the WE is converted to an output voltage signal via the trans-impedance amplifier whose gain can be adjusted to control the measured current sensitivity and the overall range. The DC level of the output voltage signal is then adjusted such that the signal spans only in the positive voltage domain that is compatible with the ADC circuit. The integrated 16 bit ADC on-board the MCU operates in a differential configuration, hence the single-ended output voltage signal is converted to a differential signal before being recorded by the ADC in the final stage of the measurement circuit. 

The complete sensor circuit system is operated by a computer which triggers the measurement, and collects the recorded DPV response data. The obtained DPV graph (\emph{applied potential versus measured current}) is then analysed using the proposed procedure as described in the following sub-section. 

\subsection{Sensor data analysis} \label{sec:analysis}

\begin{figure}[h]
\centering
\includegraphics[width=0.48\textwidth]{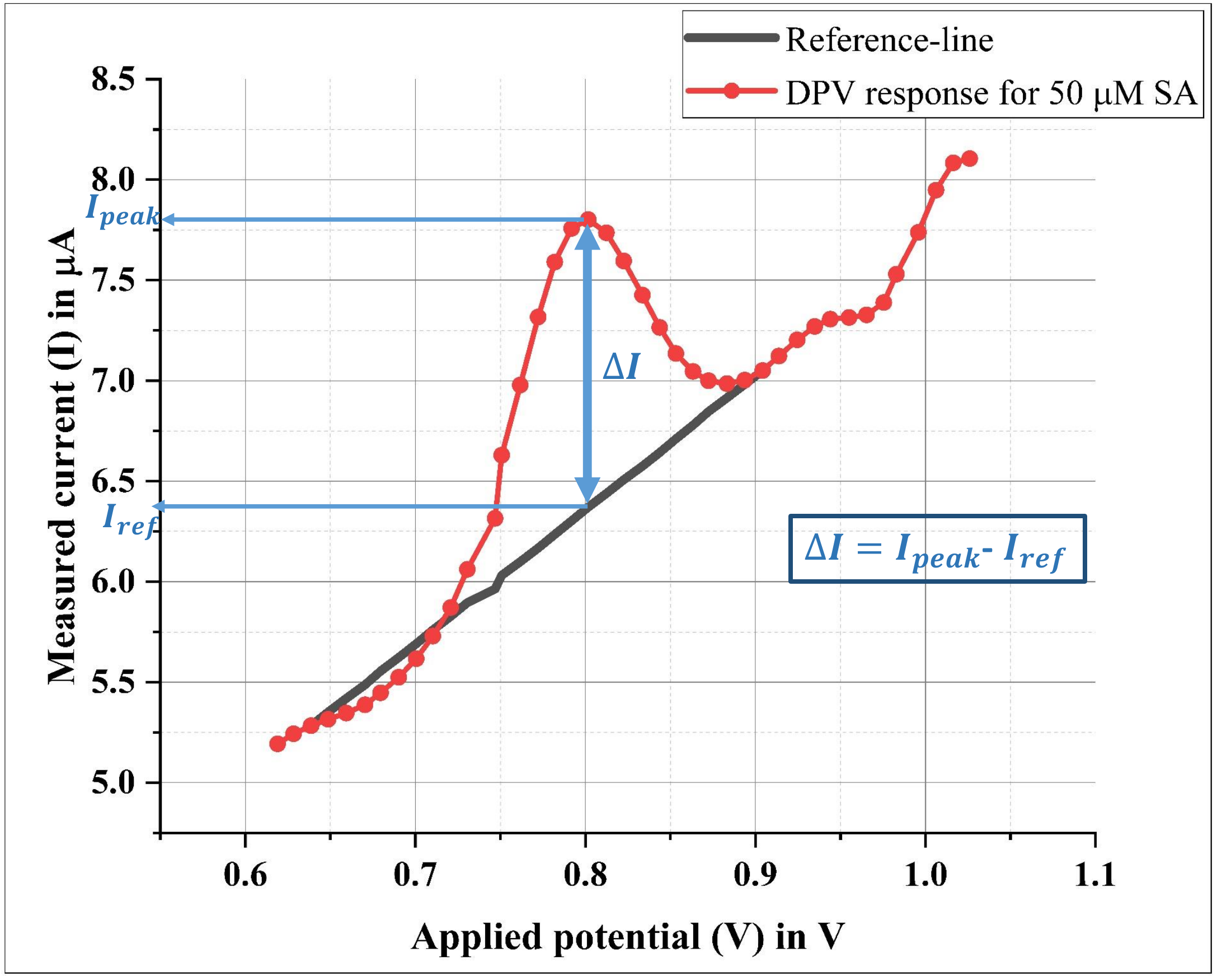} 
\caption{SA sensor data analysis method using experimentally obtained DPV graph}
\label{analysis}
\end{figure}

After the experimental DPV graph is recorded using the developed portable EC sensing system, a data extraction algorithm is implemented to obtain functional information about the SA concentration. The algorithm operates by detecting and quantifying SA specific current response within the recorded DPV graph as shown in figure \ref{analysis}. The key steps in the automated SA specific measured current-response detection algorithm are as follows: (i) SA-specific electro-oxidation current-peak detection, (ii) Determination of the reference-line, and (iii) estimation of the measured current difference ($\Delta I$). 

During the first step, a current peak ($I_{peak}$) detection procedure is implemented given by,
\begin{align}
    V_{peak} & = min[mag(\frac{dI(V)}{dV})], \\ \label{eq_main}
    & given: 0.75 < V < 0.85, \: and \nonumber \\  
    & \frac{d^2I(V)}{dV^2} < 0. \nonumber \\
        I_{peak} & = I(V_{peak})
\end{align} 

where I and V represent the current and potential variables, respectively, and V$_{peak}$ is the potential at which the current maximum occurs. Second, the initial and final potentials/currents are estimated as follows, to form the reference line:
\begin{align}
    V_{initial} & = min[mag(\frac{d^2I(V)}{dV^2})], \\
    & given: 0.6 < V < 0.75. \nonumber \\ 
    I_{initial} & = I(V_{initial}) \\
    V_{final} & = min[mag(\frac{d^2I(V)}{dV^2})], \\
    & given: 0.85 < V < 0.95. \nonumber \\
    I_{final} & = I(V_{final})
\end{align}

where (V$_{initial}$, I$_{initial}$) and (V$_{final}$, I$_{final}$) represent to two end points of the reference line. Next, using the calculated end points (current, voltage), an equation of the reference line is formed which is then used to obtain the value of I$_{ref}$ at V$_{peak}$. Finally, $\Delta I$ was calculated as the difference between I$_{peak}-$ I$_{ref}$ (as shown in figure \ref{analysis}) as the parameter of interest for determining the measured SA concentration. 

The proposed data extraction algorithm provides selectivity for SA determination in two ways: (i) The current peak in a given DPV graph occurs at the SA-oxidation reaction-specific potential, and (ii) The background current level may shift depending on the conductivity of the test solution however, the proposed procedure for determining $\Delta i$ as the ultimate measured current, this effect is nullified. Moreover, in the subsequent section, a general interference study is performed along with real-sample testing to access the applicability of the developed plug-and-play type SA-sensing system.

\section{Results and discussion} \label{result}

This section presents the results and discussion where the validation of the developed EC sensors, SA sensing and calibration, interference testing, detecting SA levels in real plant samples, and a comparative study with other recent SA sensing works are reported. 

\subsection{Portable sensing system validation}

The electrochemical operation of the developed portable EC sensing system was validated by characterizing a well-known redox-reaction involving reversible conversion between K$_3$[Fe(CN)$_6$] and K$_4$[Fe(CN)$_6$] given by,

\begin{center}
K$^+$ + [Fe(CN)$_6$]$^{-3}$ + e$^{-}$ $\Longleftrightarrow$ K$^+$ + [Fe(CN)$_6$]$^{-4}$
\end{center}

\begin{figure}[h]
\centering
\includegraphics[width=0.48\textwidth]{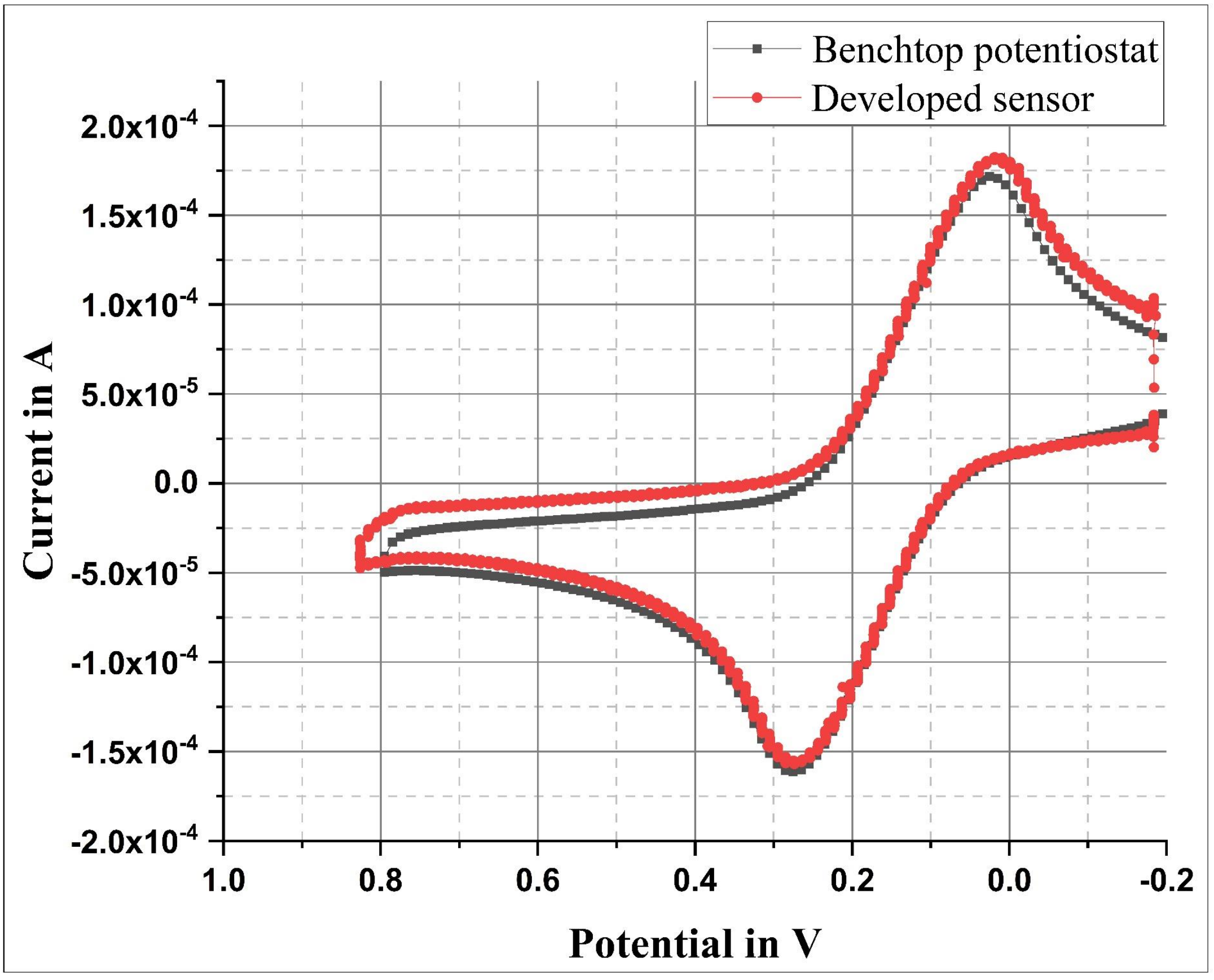} 
\caption{Comparison between the CV plots obtained using the bench-top potentiostat and the developed portable EC sensing system}
\label{valid}
\end{figure}

Figure \ref{valid} presents a comparison between the the CV response obtained using a benchtop potentiostat (CHI 660E; CHI Instruments Inc., Austin, Texas, USA) and the developed low-cost EC sensing system. It can be observed (refer Figure~\ref{valid}) that the developed plug-and-play type portable EC sensor exhibits excellent response which correlates well with the benchtop potentiostat thus validating the correct operation of the developed EC senisng system. 10 mM K$_3$[Fe(CN)$_6$] was prepared using 0.1 M KCl aqueous (DI water as solvent) electrolyte solution, and the redox reaction was carried on the disposable screen printed electrode. The following parameters were used while recording the CV responses: starting potential = -0.2 V, final potential = 0.8 V, scan rate = 0.1 V/s, and sampling rate = 100 Hz. 

\subsection{SA sensor response and calibration}

The developed plug-and-play type SA sensing system was used to obtain the DPV response and generate a calibration curve. In this section, the experimental data and the performance of the system are reported.

\begin{figure}[h]
\centering
\includegraphics[width=0.48\textwidth]{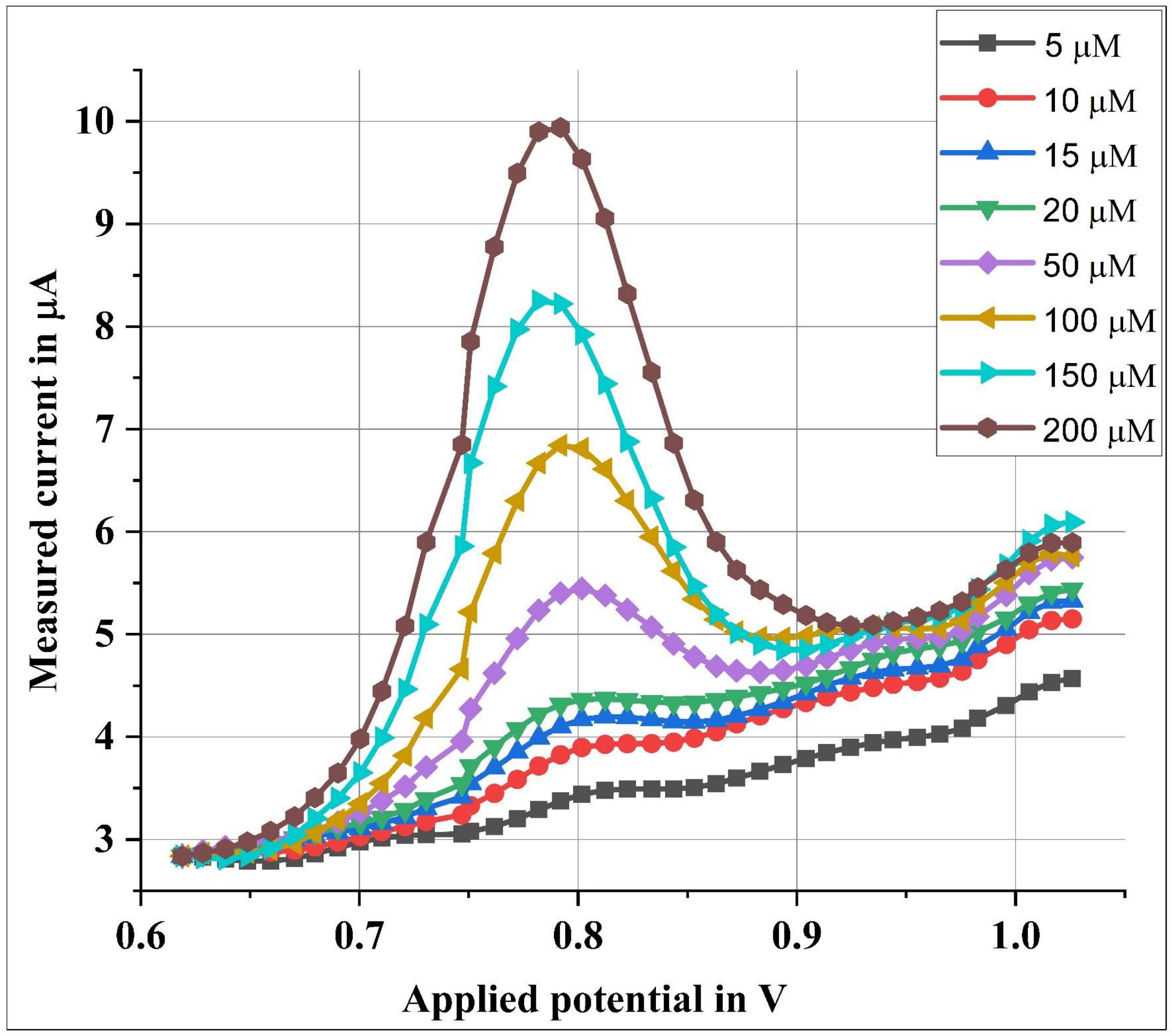} 
\caption{Experimental DPV response obtained for different SA concentrations using the developed plug-and-play SA sensing system}
\label{DPV}
\end{figure}

\begin{figure}[h]
\centering
\includegraphics[width=0.48\textwidth]{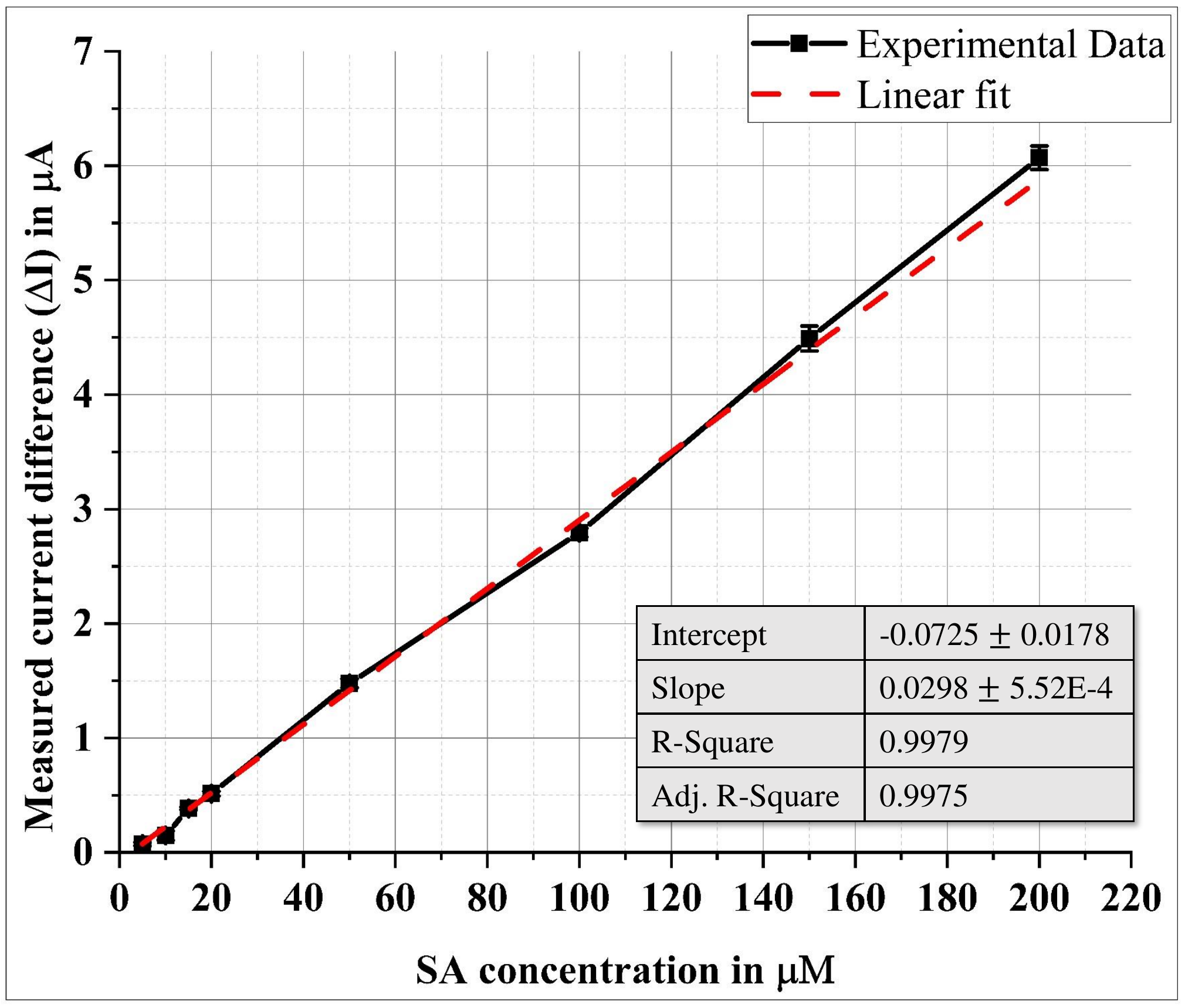} 
\caption{Measured current versus SA concentration; Linear fitting model and parameters}
\label{calib}
\end{figure}

Figure \ref{DPV} presents the experimentally obtained DPV responses for SA solutions with concentraions ranging from 5 $\mu$M to 200 $\mu$M. The experimental DPV scans were recorded with the following parameters: pulse width as 50 ms, pluse amplitude as 50 mV, sample width as 20 ms, pulse period as 500 ms, and potential increment as 10 mV. SA test solutions were prepared by first dissolving powdered SA in ethanol (as SA is high soluble in ethanol) to obtain a concentration of 100 mM (the stock solution). The stock SA solution was then diluted using aqueous 0.1 M KCl in 0.2 M 6.6 pH PBS buffer solution to obtain the desired SA concentrations. During testing, 60 $\mu$l of the SA test solutions were sampled by drop-casting them on to the carbon screen printed electrode using a pipette. The recorded DPV graphs were then used to generate a calibration function.

The experimental DPV graphs (shown in figure \ref{DPV}) were analysed according to the procedure described in \ref{sec:analysis}. Figure \ref{calib} presents the SA-specific measured current (or $\Delta I$) versus the SA concentration along with the standard error at each measurement point. Next, a linear-fitting model was applied to estimate the calibration function. It can be observed that the developed sensor exhibits excellent linear response where an R-square (also known as the coefficient of determination) value of over 0.99, a sensitivity of 0.42 $\mu$A$\cdot\mu$M$^{-1}\cdot$cm$^{-2}$, and a limit of detection of 2.54 $\mu$M (3$\times$SD$_y$/slope) was achieved. 

\subsection{Interference study and real-world testing}

\begin{figure}[h]
\centering
\includegraphics[width=0.48\textwidth]{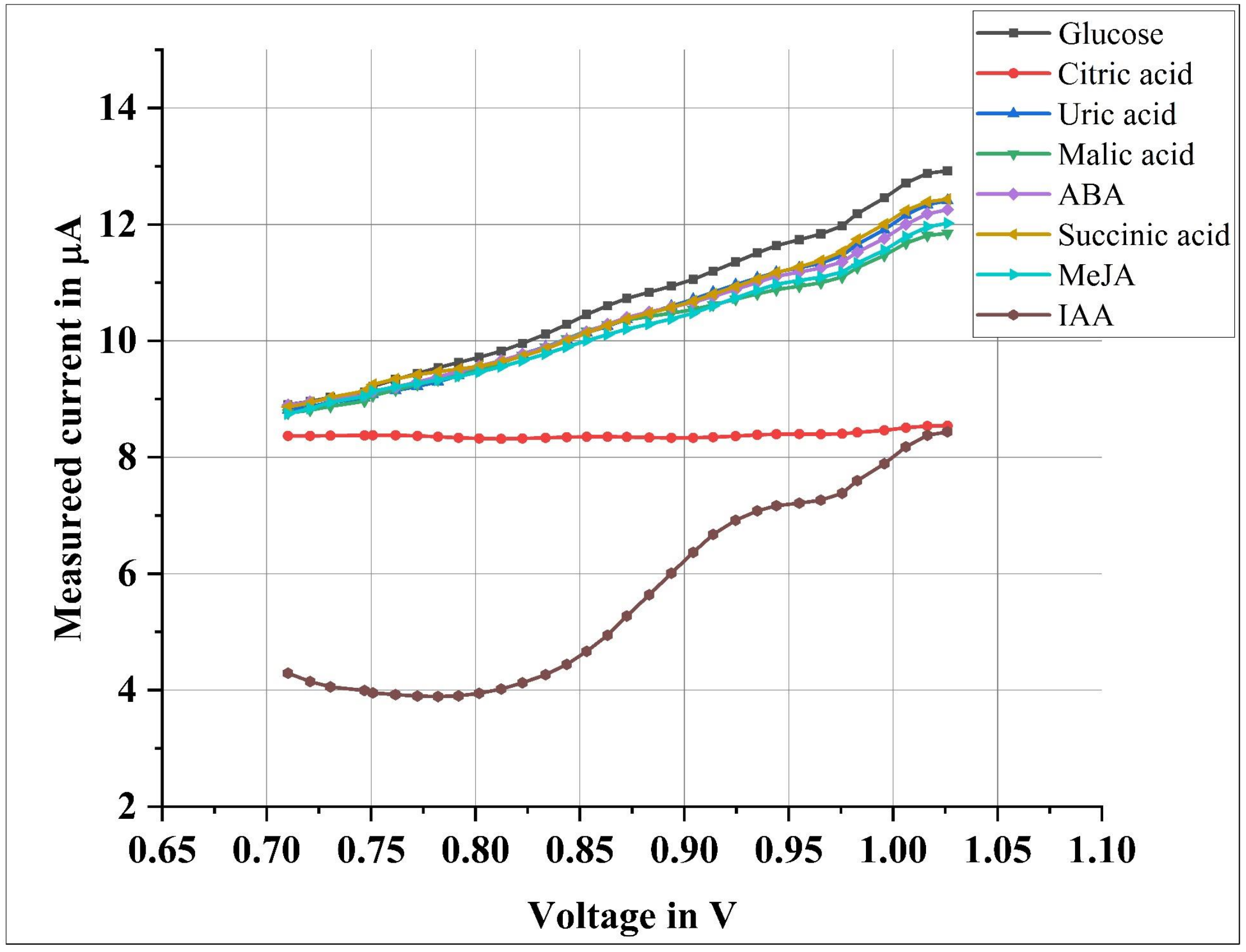} 
\caption{The DPV responses obtained from interference study; Each sample solution was tested at 0.1 mM concentration level.}
\label{inter}
\end{figure}

\begin{table}[htbp]
    \centering
    \caption{SA level testing in real samples} \label{tab:real}
    \begin{tabular}{|p{1cm}|p{2cm}|p{2.2cm}|p{1.2cm}|}\hline
        Test sample & SA level measured using the developed sensor & SA level measured using SA-UV test-kit/spectro-photometer & Percentage error (w.r.t. SA-UV kit) \\ \hline
         
         Fresh orange juice & 45.2 $\mu$M & 42.0 $\mu$M & 7.6 \% \\ \hline
         
         Fresh tomato juice & 25.4 $\mu$M & 22.8 $\mu$M & 11.4 \% \\ \hline
          
    \end{tabular}
\end{table}

\begin{table*}[htbp]
\caption{Comparison between key recent SA sensors and our work}\label{tab:comp}
\centering
\begin{tabular}{|p{2.5cm}|p{2.0cm}|p{1.5cm}|p{1.0cm}|p{3.2cm}|p{3.2cm}|p{0.6cm}|}
\hline
\textbf{SA sensor description}  &   \textbf{Sensitivity in $\mu$A$\cdot\mu$M$^{-1}\cdot$cm$^{-2}$}	& \textbf{Range in $\mu$M} &  \textbf{LOD}  &\textbf{Strengths}	& \textbf{Limitations} & \textbf{Ref., year}\\
\hline

MIP-based biosensor	& - & 0.1 to 50 & 3.9$\times$10$^{–8}$ M & Good selectivity and LOD & Complex fabrication, moderate range & \cite{SA_sens11}, 2020  \\ \hline

Bi-enzyme microfluidic biosensor & 	34.4 $\mu$A$\cdot\mu$M$^{-1}\cdot$cm$^{-2}$ & 0.5 to 64 & - &  Excellent selectivity, microfluidic device & Complex electrode fabrication, limited self-lif. & \cite{SA_sens8}, 2018  \\ \hline

Bi-enzymatic EC sensor    &  30.6 $\mu$A$\cdot\mu$M$^{-1}\cdot$cm$^{-2}$  &	 2.3 to 46.3  & 13 $nM$	& Excellent selectivity & Complex and expensive electrode fabrication, limited self-life  & \cite{SA_sens10}, 2016 \\ \hline

MIP-based biosensor	& - & 7.0$\times$10$^{–5}$ to 1.0$\times$10$^{–2}$ & 2$\times$10$^{–13}$ M & Good specificity and LOD & Limited range of operation & \cite{SA_sens7}, 2019  \\ \hline

Aptamer-based optical biosensor   &  -	&  $\sim$0.1 to 50  & 0.1 $\mu$M & Good accuracy and LOD	& Requires complex equipment and fabrication procedure, moderate selectivity & \cite{SA_sens9}, 2019\\ \hline

Electro-catalytic oxidation-based EC sensor  &   1.0 $\mu$A$\cdot\mu$M$^{-1}\cdot$cm$^{-2}$	&  5 to 1000  &   1.1 $\mu$M &  Good range of operation  & Expensive and complex fabrication method & \cite{SA_role9}, 2018 \\ \hline

Carbon nanotube/iron oxide voltammetric sensor  &   0.64 $\mu$A$\cdot\mu$M$^{-1}\cdot$cm$^{-2}$	&  0.6 to 46.3  &   0.02 $\mu$M & Good LOD & Limited range, selectivity not tested & \cite{SA_sens13}, 2019 \\ \hline

Screen printed electrode-based EC sensor   & 1.2 $n$A/$\mu$M  &   16 to 300   &   5.6 $\mu$M   & Indirect determination of SA was proposed for improved specificity & Limited range, poor sensitivity, and moderate LOD & \cite{SA_sens14}, 2018  \\ \hline

In-vivo EC detection of SA in seedlings   & -  &   0.1 to 1   &   48 $p$M   &  In-vivo sensing was demonstrated, complex electrode fabrication	&   Limited range of operation  & \cite{SA_sens18}, 2018  \\ \hline

Carbon fiber electrode-based EC sensor   &  0.21 $n$A/$\mu$M &   2 to 3000   &   1.7 $\mu$M   & Good linearity due to electrode cleaning with NaOH to remove secondary SA products & Poor sensitivity, selectivity was not reported & \cite{SA_sens16}, 2016  \\ \hline

A portable plug-and-play type EC SA sensor   &  30 $n$A/$\mu$M or 0.42 $\mu$A$\cdot\mu$M$^{-1}\cdot$cm$^{-2}$ 	&   5 to 200   &   2.5 $\mu$M   & Excellent sensitivity, selectivity and linearity, portable, low-cost, easy to use, ideal for in-field testing	&  Moderate LOD  &   our work \\

\hline
\end{tabular}
\end{table*}

In order to access the real-world applicability of the developed SA sensing system, an interference study was performed where current response due to other species that may be present in agricultural samples are observed. The interfering responses of the following chemical species were recorded: glucose, citric acid, uric acid, malic acid, abscisic acid (ABA), succinic acid, methyl-jasmonate (MeJA) and indole-3-acetic acid (IAA). These biochemicals were selected based on prior studies reported in literature \cite{SA_sens18, SA_role9, SA_sens15}.   

Figure \ref{inter} presents the DPV responses recorded for key known compounds that may be present in the plant/agricultural samples. It can be observed that there is no detectable peak in the potential range of interest as determined in \eqref{eq_main}, corresponding to SA-specific signal. Moreover, as described in the aforementioned section, the proposed data analysis procedure is independent of the base/background current level variations that may result for variable ionic composition of the test sample. During the interference, the samples were prepared using the same buffer that was used to prepare SA samples, that is aqueous solution of 0.1 M KCl in 0.2 M ph 6.6 PBS buffer. 

Table \ref{tab:real} presents the results obtained by measuring SA levels in real plant/fruit samples including fresh orange and tomato juice. For reference or ground-truth values of SA concentration in the real test samples, a commercially available enzyme kit (Salicylate liqui-UV test kit; refer section \ref{material}) was used. The working principle is based on the enzymatic reaction,

\begin{center}
SA + NADH + O$_2$ + 2H$^{+}$ $\xrightarrow{\text{SH}}$ Catechol + NAD$^+$ +  \\
\hspace{5.5cm} H$_2$O + CO${_2}$
\end{center}

where the amount of Nicotinamide adenine dinucleotide (NADH) consumed is proportional to the concentration of SA in the solution. The measurement procedure involved the use of a spectrophotometer, where the light absorption through the solution is measured at 320 nm (absorption wavelength for NADH) of reference versus the test solution where the change in light absorption intensity is recorded. The method was calibrated by first measuring the intensity change for a known SA concentration solution followed by real sample measurements.

\subsection{Comparison with related works}
A detailed comparison between the key recent SA sensors reported in literature including our work is presented in Table \ref{tab:comp}.

\section{Conclusion}

A first-of-its-kind plug-and-play-type SA sensing system was reported in this work. The developed system consisted of the following key components: a 3D-printed electrode adapter for electrode-sensor interface, the specially designed sensing electronics, and the sensor data analysis procedure to extract functional SA level information. The proposed SA sensing system exhibited excellent linearity and performance where a sensitivity of , and limit of detection of was observed. Additionally, an interference study was performed to access the practical applicability of the developed system, and SA concentrations in real samples were estimated with about 90\% accuracy. The overall cost of the developed system for a single unit was around \$50 making it ideal for various applications spanning agriculture, healthcare and pharmaceutical. 

Following on the development of the proposed sensor in this work, the future work may focus on integrating a battery and a Bluetooth module with the MCU for wireless data recording and analysis. Secondly, the electrode may be functionalized with nano-materials to improve sensitivity even further but may result in increased cost and complex sensor fabrication procedure. Finally, the developed system can be used for more extensive studies for monitoring SA levels in Agricultural and pharmaceutical sensing applications

\appendices
\section{Schematic of P-AFE with component details}

\begin{figure}[h]
\centering
\includegraphics[width=0.49\textwidth]{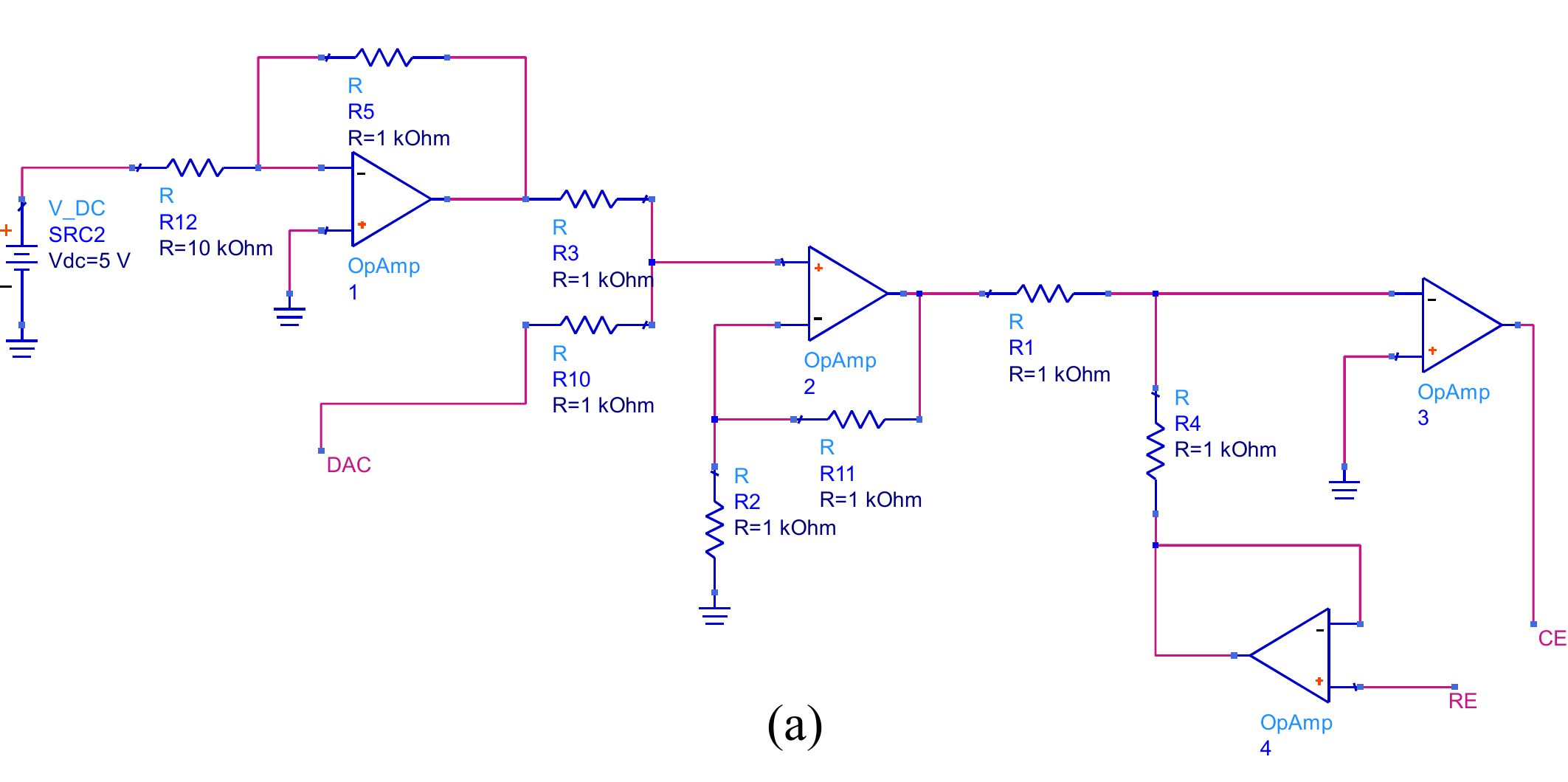} 
\end{figure}
\begin{figure}[h]
\centering
\includegraphics[width=0.49\textwidth]{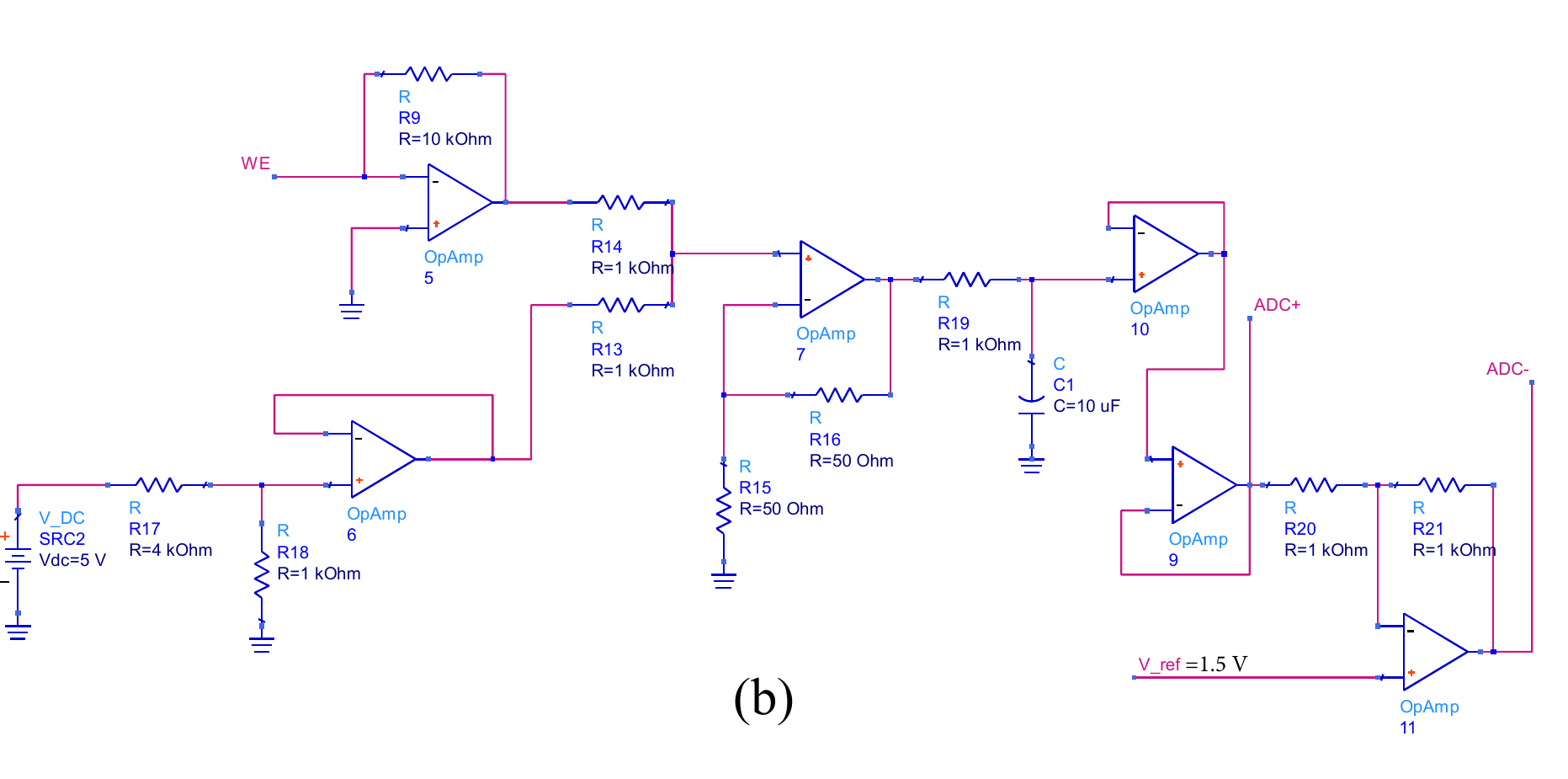} 
\caption{A complete schematic of the developed P-AFE including the component values used. For displaying purposes the overall circuit design was divided in to two parts: (a) and (b). DAC, ADC+ and ADC- were connected to their respective pins on the MCU (C2000 Launchpad)}
\label{Fig:appen}
\end{figure}

\begin{figure}[htbp]
\centering
\includegraphics[width=0.31\textwidth]{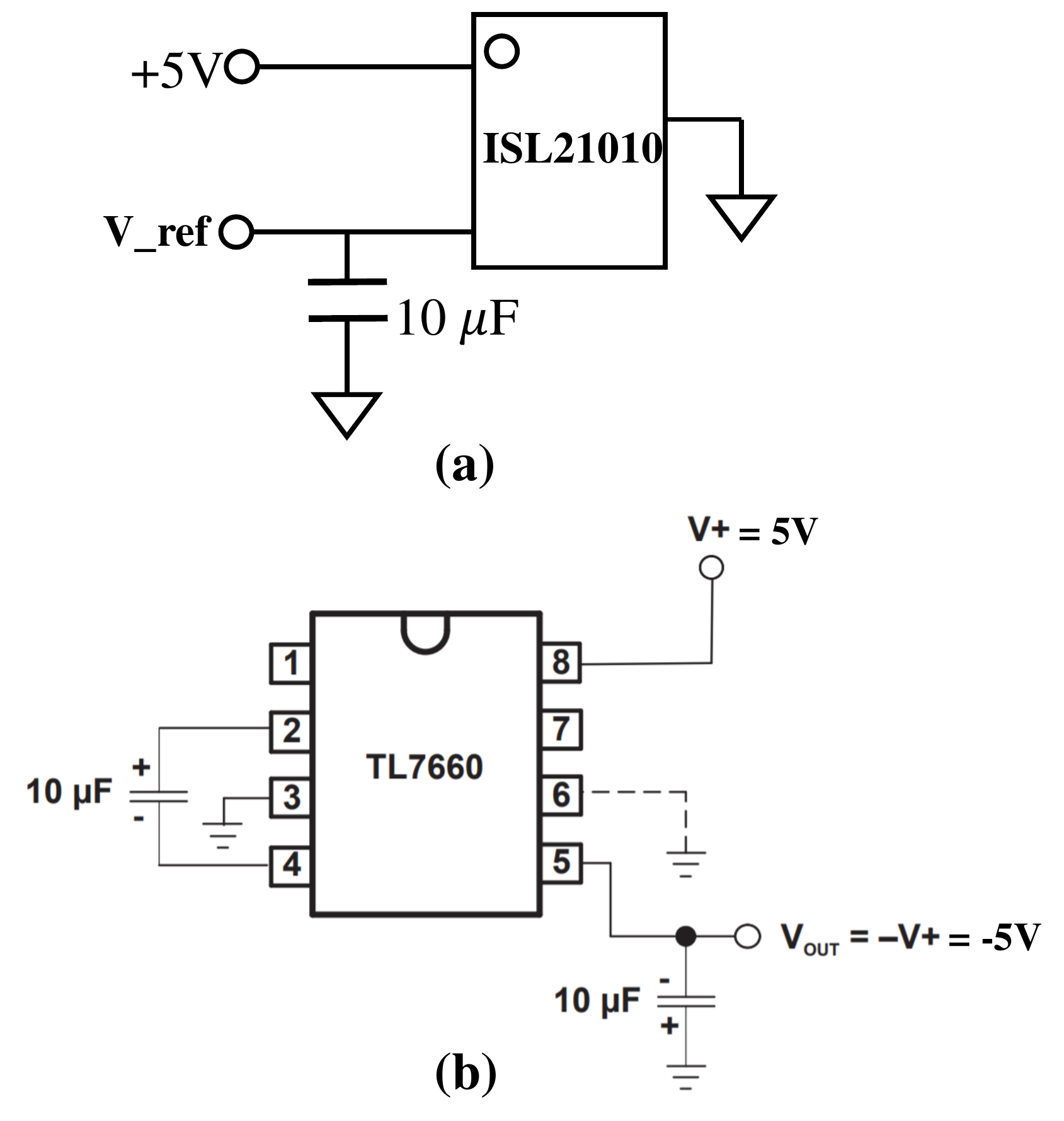} 
\caption{Circuit designs for: (a) ISL21010 (V\_ref generator), and (b) TL7660 (voltage converter) \cite{app_1}}
\label{Fig:IC}
\end{figure}
Figure \ref{Fig:appen} shows the complete schematic of the developed P-AFE circuit. The 5 V on-board power supply from the MCU was used to power the P-AFE. The following components were used in the P-AFE design: (i) All the operational amplifiers (OPAMPs) were obtained from the LM324AN integrated circuits (ICs), (ii) ISL21010 IC was used to generate the V$\_$ref voltage (1.5 V), and (iii) Voltage converter TL7660 IC was used to generate the -5 V supply for powering the OPAMPs. Figure \ref{Fig:IC} shows the circuit design for the ICs: ISL21020 and TL7660. 


\bibliographystyle{unsrt}
\bibliography{main}


\begin{IEEEbiography}[{\includegraphics[width=1in,height=1.25in,clip,keepaspectratio]{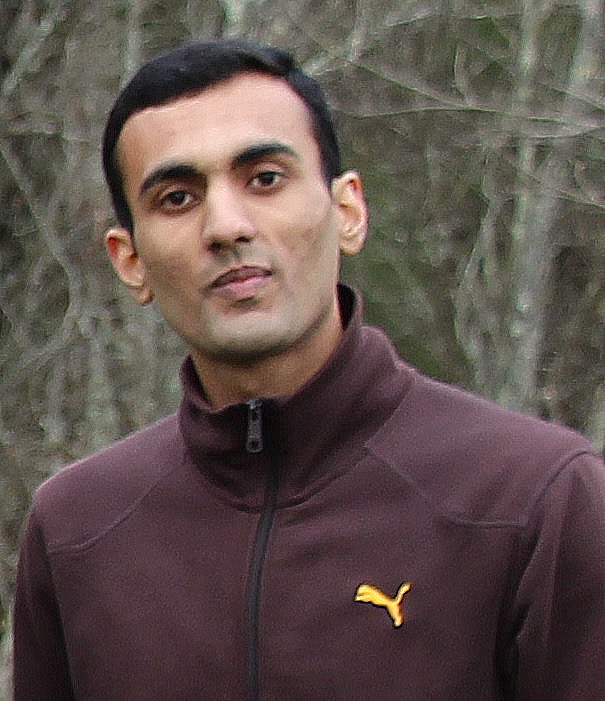}}]{Bhuwan Kashyap} received the B.Tech. degree in electronics and communication engineering from the Indian Institute of Technology Roorkee, India, in 2014. He is currently pursuing the Ph.D. degree in electrical engineering with Iowa State University, Ames, IA, USA.
His current research interests include designing sensing technologies for sustainable agriculture, microwave engineering, micro-electro-mechanical systems, digital circuit design, lab-on-a-chip-based systems and clean energy harvesting devices.
\end{IEEEbiography}

\begin{IEEEbiography}[{\includegraphics[width=1in,height=1.25in,clip,keepaspectratio]{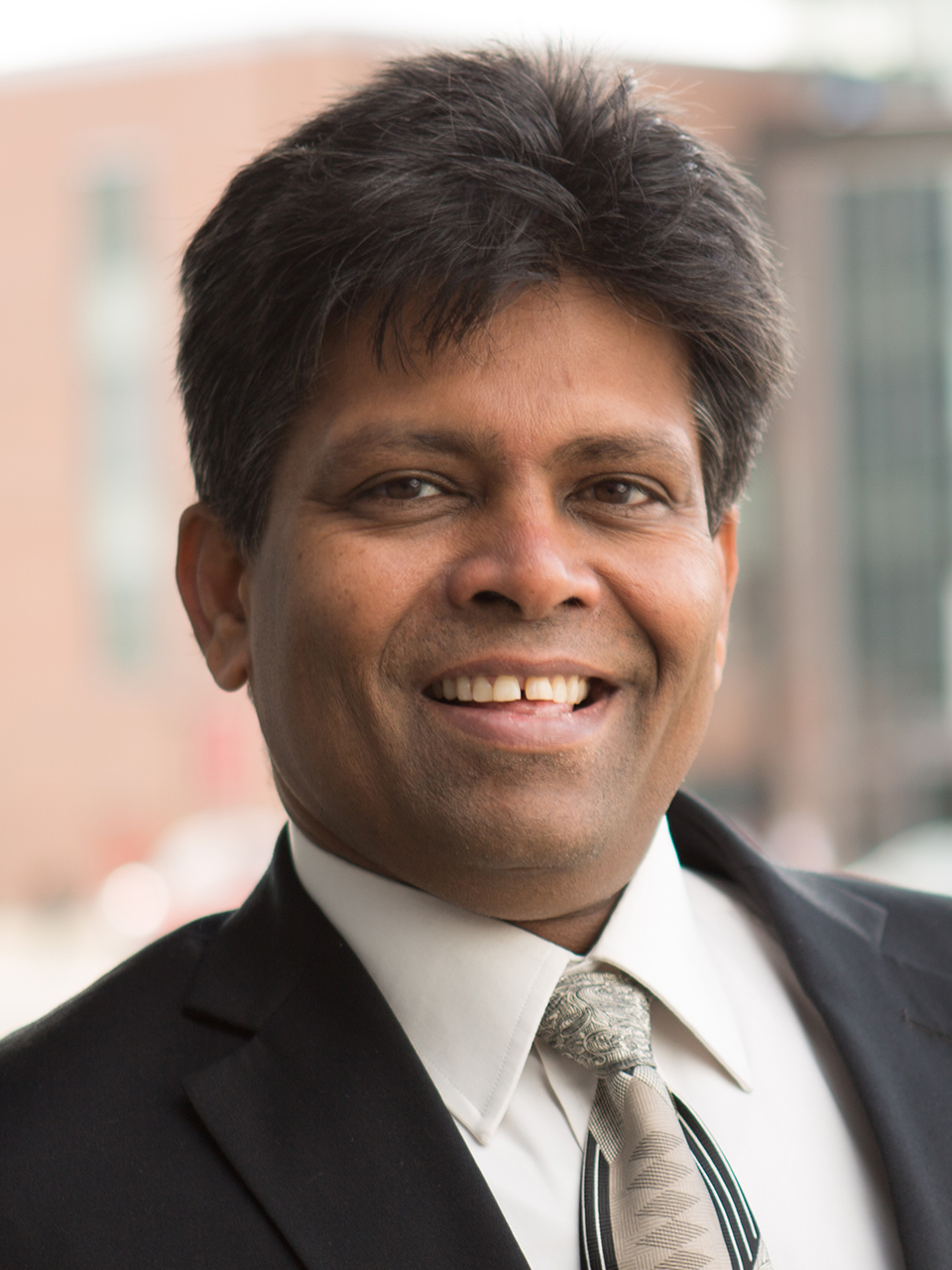}}]{Ratnesh Kumar} (F’07) is a Harpole Professor at the Iowa State University, Electrical and Computer Engineering, where he directs the ESSeNCE (Embedded Software, Sensors, Networks, Cyberphysical, and Energy) Lab. Previously, he held faulty position at the University of Kentucky, and various visiting positions with the University of Maryland (College Park), the Applied Research Laboratory at the Pennsylvania State University (State College), the NASA Ames, the Idaho National Laboratory, the United Technologies Research Center, and the Air Force Research Laboratory. He received a B. Tech. degree in electrical engineering from IIT Kanpur, India, in 1987 and the M.S. and Ph.D. degrees in electrical and computer engineering from The University of Texas at Austin in 1989 and 1991 respectively. Ratnesh was a recipient of the Gold Medals for the Best EE Undergrad, the Best EE Project, and the Best All Rounder from IIT Kanpur, the Best Dissertation Award from UT Austin, the Best Paper Award from the IEEE Transactions on Automation Science and Engineering, and Keynote Speaker and paper awards recipient from multiple conferences. He is or has been an editor of several journals (including of IEEE, SIAM, ACM, Springer, IET, MDPI), was a Distinguished Lecturer of the IEEE Control Systems Society, is a recipient of D. R. Boylan Eminent Faculty Award for Research from Iowa State University, a Fellow of IEEE, and also a Fellow of AAAS.
\end{IEEEbiography}

\end{document}